\documentclass[12pt,prd,aps,tightenlines]{revtex4-1}
\pdfoutput=1 
\renewcommand{\thefootnote}{\fnsymbol{footnote}}
\usepackage{amssymb}
\usepackage{amsbsy}
\usepackage{chngcntr}
\usepackage{amsmath}
\usepackage{amsfonts}
\usepackage{verbatim}
\usepackage{graphicx}

\renewcommand{\(}{\left(}
\renewcommand{\)}{\right)}

\def\roughly#1{\mathrel{\raise.3ex
\hbox{$#1$\kern-.75em\lower1ex\hbox{$\sim$}}}}

\newcommand{\be}{\begin{equation}}

\newcommand{\ee}{\end{equation}}
\newcommand{\bqa}{\begin{eqnarray}}
\newcommand{\eqa}{\end{eqnarray}}

\begin{document}

\rightline{BI-TP 2015/10}

\rightline{TUW-15-22}

\vspace{1.0cm}

\title{Comment on Hawking radiation and trapping horizons}

\author{ R.~Baier} 
\email {baier@physik.uni-bielefeld.de}

\affiliation{Faculty of Physics, University of 
Bielefeld, D-33501 Bielefeld, Germany}

\author{ S. A. Stricker} 
\email {stricker@hep.itp.tuwien.ac.at}
\affiliation{Institute for Theoretical Physics, Vienna University of Technology, Wiedner Hauptstr. 8-10, A-1040 Vienna, Austria}

\begin{abstract}
\vspace{2cm}
We consider dynamical black hole formation from a collapsing fluid described by a symmetric and flat FRW metric. 
 Using the Hamilton-Jacobi method the local Hawking temperature for the formed  trapping/apparent horizon is calculated. 
The local Hawking temperature depends on the tunneling path, which we take to be along a null direction $(\Delta s=0)$.
We find that the  local Hawking temperature depends directly  on the equation of state of the collapsing fluid.
We argue that  Hawking radiation by quantum tunnelling from  future inner and future outer trapping horizons is possible. 
However, only radiation from a  space-like dynamical  horizon has a chance to be observed by an external observer.
Some comparison to existing  literature is  made. 

\end{abstract}

\maketitle

\section{Introduction}

There exists a huge amount of literature  investigating  Hawking radiation in various analytical models of black hole formation. 
In time dependent backgrounds an important concept are trapping  horizons (surfaces) which can be classified in various subcategories. 
In this context one interesting question to ask is: For which trapping horizons can Hawking radiation appear and is there a preferred path?
In this short note we examine  this question  for  dynamical black hole formation from  a collapsing fluid described by a symmetric and flat $(k=0)$ Friedman-Robertson Walker (FRW) metric.
Using the Hamilton-Jacobi (HJ)  tunneling method we calculate the  local Hawking temperature  for the formed apparent horizon (which can be a   future inner or a future   outer trapping horizon) with radius
$R$ and mass $m(t,R)$.

 For tunneling along a null direction $\Delta s = 0$
 we argue that in summary  the absolute value, evaluated on the horizon $R_H = 2 m$,  has to be taken, 
\be
T_H 
    = \frac{\vert H + \frac{ \dot{H}}{2 H}\vert}{2 \pi } \Big|_H ~,
\ee
where $H$ is the Hubble parameter, which in terms of the scale function $a$ reads $H (t)= \frac{\dot{a}(t)}{a(t)}$.
In terms of the surface gravity $\kappa_H$ on the dynamical horizon the temperature is
\be
T_H = \frac{\vert \kappa_H \vert}{2 \pi}.
\ee
As a consequence, by looking at the temperature alone computed by the HJ method  one can not decide if Hawking radiation for a certain type of apparent horizon is present. 
Note that in most of the literature in which the tunnelling method is applied to obtain the Hawking temperature for dynamical black holes no absolute values are invoked, including
 \cite{Parikh:1999mf,Visser:2001kq,Vanzo:2011wq,Vanzo:2008uq,Nielsen:2005af,Nielsen:2008cr,Nielsen:2008dj,Tian:2014sca,Ashtekar:2004cn,Hayward:2008jq,DiCriscienzo:2010zza,Senovilla:2014ika}.

The paper is organised as follows. In section~\ref{model} we review the collapsing fluid model after which we calculate the surface gravity in section~\ref{sec:SG}. In section~\ref{HJ} we calculate the local Hawking temperature for space like and timelike horizons, mostly following  the review paper \cite{Vanzo:2011wq}.
\section{The collapsing fluid model}\label{model}
In this section we review the most important details of the explicit example of a collapsing fluid needed for the rest of this work, where we closely  follow the notation of the paper \cite{Baier:2014ita}. For related works on collapsing fluids  see also   \cite{Joshi:2002,Joshi:2011hb,Joshi:2008zz,Joshi:2007zza} and  \cite{Hayrev}.

The spherically symmetric collapse  is expressed in terms
of the scale factor $a(t)$, with $\dot{a} (t) = \frac{da}{dt} < 0$ and 
using the FRW metric
\be
ds^2 = -dt^2 + a^2 (t) (dr^2 + r^2 d \Omega^2_2).
\label{FW}
\ee
We take  the marginally bound case ($k=0$), which allows to obtain  a 
simple  analytic solution.
The scale factor $a(t)$ is determined from  Einstein's equations 
(with vanishing cosmological constant and back reaction ignored)
\be
\dot{a}^2 = \frac{1}{3} \rho a^2,
\label{7}
\ee
and
\be 
\ddot{a} = - \frac{1}{6} (\rho + 3p) a .
\label{8}
\ee
The time dependent energy density $\rho(t)$ and the pressure $p(t)$ of the
 fluid satisfy the following  equation of state (EoS) 
\be
p = w\rho = - \frac{1 + 2\beta}{3} \rho, 
\label{4}
\ee
where $\beta$ is constrained to lie in the interval $-2\le \beta <1$.
One can check that Eqs.~(\ref{7} - \ref{4}) are compatible with the ansatz
\be
\dot{a}^2 = a^{2 \beta},
\label{10}
\ee
which then gives, choosing $a(0) = 1$ as initial condition,   the solution
\be
a (t) = 
\bigg\{ 1 -  t/t_s \bigg\}^{\frac{1}{1-\beta}}.
\label{17}
\ee
The singularity formation time is given by  $t_s = \frac{1}{1-\beta}$ and $0 < t < t_s$.
For later purpose the Hubble parameter is introduced,
\be
H(t) = \frac{\dot{a}(t)}{a(t)} = - a^{\beta - 1} = - \frac{1}{1- t/t_s} \le 0~,
\label{Hubb}
\ee
which for $\beta = - \frac{1}{2}$  coincides
with the $k=0$ and $p(t)=0$ Oppenheimer-Snyder  model \cite{Oppenheimer:1939ue,Adler:2005vn}.
For  $\beta = - {2}$ it determines the stiff fluid $(p =\rho)$.

Since the  existence of trapped surfaces is crucial for the determination of the position of the apparent horizon we next  repeat  under which conditions trapped surfaces form
 \cite{Hayward:1994,Hayward:1996,Hayward:2000ca,Booth:2005qc,Booth:2005ng,Ashtekar:2004cn}.
For the FRW metric (\ref{FW}) the following out and ingoing  null geodesics are introduced (see also \cite{Poisson:2004}),
\be
 l_{\mu} = ( -1, a(t), 0, 0)~, ~~n_{\mu} = \frac{1}{2} (-1 , -a(t), 0 ,0),  \qquad  
\label{tr1}
\ee
with $l_{\mu}l^{\mu} = n_{\mu} n^{\mu} = 0$ and  $l_{\mu} n^{\mu} = -1$.
From theses geodesics  the outgoing (ingoing) null expansions $\theta_+~(\theta_-)$ are constructed as 
\be
\theta_{+} = h_{\mu \nu} \nabla^{\mu} l^{\nu},~~~ \theta_{-} = h_{\mu \nu} \nabla^{\mu} n^{\nu},
\label{d+}
\ee
where  the transverse metric 
\be
h_{\mu \nu} = g_{\mu \nu} + l_{\mu} n_{\nu} + n_{\mu} l_{\nu}~,
\label{hmu}
\ee
satisfies $h_{\mu \nu} l^{\nu} = h_{\mu \nu} n^{\nu} =0$.
Furthermore, introducing the new variable 
\be
 R(r,t) = r a(t),
\label{radR}
\ee
 which corresponds to the physical radius of the collapsing matter, one obtains for the FRW metric (\ref{FW}) 
\cite{Booth:2005ng}
\be
\theta_+ = 2 \frac{\dot{R} + 1}{R} = 2 \(\frac {\dot{a}}{a} + \frac{1}{a\,r}\)~,~~~~\theta_-=  \frac{\dot{R} - 1}{R} =  \frac {\dot{a}}{a} -  \frac{1}{a\,r}~,
\label{th+}
\ee
which leads to the expansion
\be
\theta = \theta_+ \theta_- = \frac{2}{a^2} \( {\dot{a}}^2 - \frac{1}{r^2}\)~
= ~ 2 (H^2 - 1/R^2)~.
\label{thet}
\ee
Marginally trapped surfaces have the property  $\theta = 0$,
\be
 \theta_{+} = 0,~~~  \theta_{-} < 0~,
\label{trapp}
\ee
from which  the location of the apparent horizon is obtained
\be
r_{H}= - \frac{1}{ \dot{a}(t) }.
\label{cond}
\ee
An equivalent definition for the boundary of a possible apparent horizon is given by
\be
g^{\mu\nu} \partial_\mu R\, \partial_\nu R = 0,
\label{23}
\ee
which  for the metric at hand  (\ref{FW}) reads
\be
\dot{R}^2 (r,t) = r^2 \dot{a}^2 (t) = 1.
\label{24}
\ee
The above condition can be translated into a condition on the time dependent   Misner-Sharp mass \cite{Misner:1964je} given by 
\be
2 m(r,t) = R (1-g^{\mu \nu} \partial_\mu R \partial_\nu R) = 
r^3 a (t) \dot{a}^2 = R^3 H^2~,
\label{32}
\ee
from which one deduces that marginally trapped surfaces are present whenever the condition 
\be
R(r,t)\Big|_H = R_H=2m (R_H,t)
\label{31}
\ee
can be satisfied.
If the above condition is fulfilled,   the marginally trapped surface ($\theta_+=0$) can be further classified into a future inner/outer trapping horizon given by the condition \cite{Hayward:1994}
\bqa
\rm{future~ inner}&:&~~\theta_-<0~~\&~~\partial_- \theta_+>0~,\nonumber \\ 
\rm{future~ outer}&:&~~\theta_-<0~~\&~~\partial_- \theta_+<0~,
\eqa
with $\partial_-=2n^\mu\partial_\mu=(\partial_t-\frac{1}{a} \partial_r)$ being the Lie  derivative along future directed ingoing null geodesics. Using   Einstein's equations  (\ref{7}) and (\ref{8}) and $\theta_+=0$ the condition for the apparent horizon becomes
\be
2 n^{\mu} \partial_{\mu} \theta_+ = 
(1 + \beta)  \frac{1}{r^2 a^2 } ~=~(1 + \beta)  \frac{1}{{R_H}^2 } ~,
\label{inout}
\ee
which is positive (future inner horizon)  for $ - 1 < \beta < 0$ and negative (future outer horizon)  for $\beta<-1$. The future outer trapping horizon provides a general definition of a black hole \cite{Hayward:1994}.

With the new variable $R$ introduced in (\ref{radR})  the metric becomes
a special case of the Painlev\'e-Gullstrand (PG) form \cite{Poisson:2004,Adler:2005vn}, namely
\be
ds^2 = - [1 - \psi^2 (t,R)] dt^2 + dR^2 - 2 \psi (t,R) dRdt + R^2
d \Omega^2_2\; ,
\label{2a}
\ee
where
\be
\psi (t,R) = R \frac{\dot{a}(t)}{a(t)}  =  R H < 0~.
\label{3a}
\ee
The same   Painlev\'e-Gullstrand form is used  in Ref. 
\cite{Visser:2001kq}, Eq. (2.2) as can be seen  by identifying  $r=R$ and
\be
c (t,R) = 1,~~ v (t,R) \equiv \psi (t,R).
\label{4a}
\ee

\noindent
In comparison with the notation used in \cite{Vanzo:2011wq} section 4.3.4 under
"The synchronous gauge" we note
\be
a^2(t)  = 1/B(r,t), 
\ee
 and Eq.~(\ref{radR}).

\section{Surface gravity}\label{sec:SG}
Having identified the location of the apparent horizon we can now evaluate the geometrical surface gravity for the metric (\ref{2a})  at the apparent horizon via the formula \cite{Hayward:1997jp,Hayward:2008jq}
\be
\kappa_H = \frac{1}{2 \sqrt{-\gamma}} 
\partial_i~(\sqrt{-\gamma} \gamma^{ij}~\partial_j R) \Big|_H~, ~ i,j = t,R~,
\label{SG}
\ee
where $\gamma$ is the 2 dimensional submanifold  spanned by the $t$ and $R$ coordinates. For the metric (\ref{2a}) its components are given by  $\gamma_{tt} = -(1-\psi^2), \gamma_{tR} = \gamma_{Rt} = -\psi, \gamma_{RR} =1$,
leading to the surface gravity 
\be
\kappa_H = - \partial_t \psi/2 + \partial_R \psi =  - \dot{\psi}/2 + {\psi}^\prime = H + \frac{\dot{H}}{2 H} ~,
\label{kapp}
\ee
which can also  be written as
\be 
\kappa_H = - \frac{1 + \beta}{2 R_H} =
 - [n^{\nu}\partial_{\nu} \theta_+]\Big|_H~ R_H~,
\label{kappa2}
\ee
where in the last equality we used (\ref{inout}). For comparison with existing literature it is convenient to express  the surface gravity  in terms of the Misner-Sharp mass (\ref{32}). Using  the relation 
\be
\psi = - \sqrt{\frac{ 2 m(t,R)}{R}},
\label{Msharp}
\ee
together with (\ref{kapp}) one obtains
\be
\kappa_H = \frac{ 1 - 2 {m}^\prime + \dot{m}}{2 R_H} \Big|_H ~,
\label{kapp3} 
\ee
which is consistent with results presented  in  \cite{Senovilla:2014ika,Pielahn:2011ra}.

To be more explicit, for the $k = 0$ fluid the mass in PG coordinates is given  by
\be
2m = R^3 \left( \frac{\dot{a}}{a}  \right)^2 = R^3 H^2.
\ee

At this point it is important to note that the surface gravity $\kappa_H$ is only positive for future outer trapping horizons \cite{Hayward:1994,Hayward:1996},
i.e. for $\beta < -1$ or equivalently $\omega > 1/3 $.

\section{Hamilton-Jacobi tunneling}\label{HJ}
We are now  going to derive the local Hawking temperature of apparent horizons using the Hamilton-Jacobi tunnelling method, following mainly
  \cite{Vanzo:2011wq}, sections 2.3, 4.3.3 and 4.3.4 (but see also 
\cite{DiCriscienzo:2010zza,Vanzo:2008uq,Visser:2001kq,Nielsen:2005af,Nielsen:2008cr,Nielsen:2008dj,Cai:2008gw}).
This is done by considering the Hamilton-Jacobi (HJ) equation for 
the action on a curved background,
\be
 g^{ij} \partial_i I \partial_j I = 0~,~i,j = t, r
\label{7b}
\ee
and
\be
I = \int_{\gamma} dx^i \partial_{i}I ~,
\label{7a}
\ee
where the integration has to be taken  along a certain oriented path $\gamma$ \cite{Hayward:2008jq,Vanzo:2011wq}.
One can then obtain the radiation temperature $T_H$ from the definition of the tunnelling rate in terms of the imaginary part of the action
\be
\Gamma \equiv e^{-2 \Im m {I}}  \equiv e^{-\omega / T_H}.
\label{17a}
\ee

\noindent
Using the metric Eq.~(\ref{FW}) the HJ equation reads

\be
\partial_t I = -\frac{1}{a} \partial_r I~.
\label{HJe}
\ee
Following the null path $\gamma$ for which  $\Delta s =0$, i.e.
 $\Delta r = - \frac{\Delta t}{a(t)}$,
one obtains
\be
 I = 2 \int_{\gamma} \partial_r I dr~.
\ee
Note that for   $\Delta r =  + \frac{\Delta t}{a(t)}$ one obtains $I = 0$.
Therefore either positive $\Delta r$ and negative $\Delta t$
or negative  $\Delta r$ and positive  $\Delta t$ must be chosen.

\noindent
Using the Kodama energy vector
\be
K^i = \frac{1}{\sqrt{-\gamma}} \epsilon^{ij} \partial_j R(t,r),
\ee
i.e $K^t = 1, ~K^r = -r H$, and the invariant energy
\be
\omega = -K^i \partial_i I = \(\frac{1}{a} + rH \) \partial_r I~,
\ee
or equivalently 
\be
 \partial_r I = \frac{a \omega}{1 +a r H}~= \frac{a \omega}{1 + RH}~,
\ee
the imaginary part of the action can be written as
\be\label{INTE}
\Im m I = 2~\Im m \int_{\gamma} \frac{a \omega}{1 + a r H(t)} dr~.
\ee
In the following we are interested in the near-horizon approximation 
$ r \approx r_H =  - \frac{1}{\dot a}$,  i.e. $R_H H = -1$, for which the denominator of (\ref{INTE}) becomes
\be\label{nearhorizon}
1 + a r H \approx 2 a (H + \frac{\dot H}{2 H})~ (r-r_H) = 2 a \kappa_H (r - r_H)~.
\ee

\subsection{Space-like horizon}
We can now calculate the local temperature of the black hole in the presence  of space-like horizons. 
Using the Feynman $i\epsilon$ prescription after the near horizon  expansion
and taking $\kappa_H$  out of  the integral in
Eq.~(\ref{INTE}) one obtains
\be
\Im  m I = \frac{1}{\kappa_H} 
 \Im m  \int_{\gamma} \frac{\omega}{r - r_H -i\epsilon} dr =
\frac{\pi \omega}{\kappa_H} ~,
\label{intsea}
\ee
in terms of the surface gravity. Note that the path $\gamma$ is taken
 with increasing radial coordinate
$\Delta r > 0$, which  is denoted by $\int_{\gamma} =
 {\int \hspace*{-0.5cm} \searrow}$ in \cite{Vanzo:2011wq}.

\begin{figure}
\begin{center}
\includegraphics[scale=0.6]{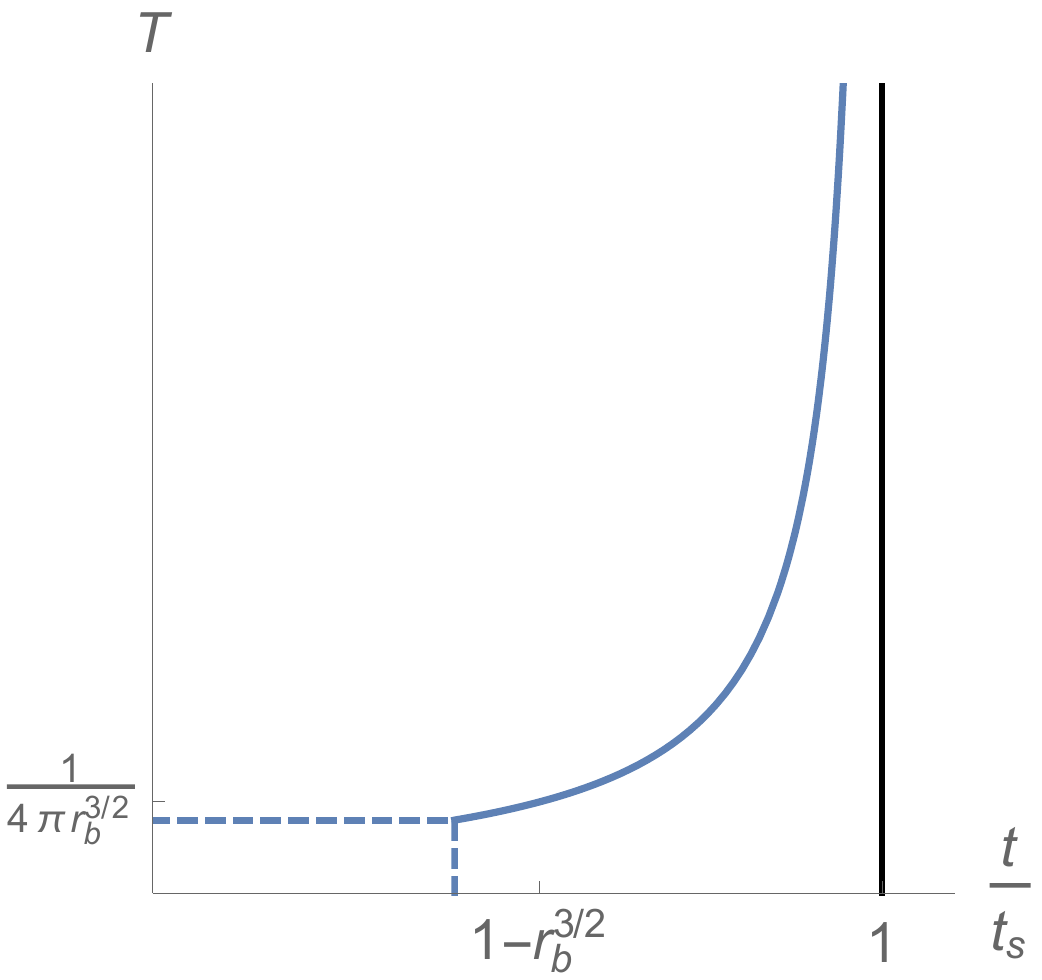}
\caption{\label{temper} Hawking temperature (blue) as a function of time in units of $t_s$ for the stiff fluid case $\rho=p$.}
\end{center}
\end{figure}

The Hawking temperature 
\be
 T= \frac{\kappa_H}{2 \pi}
\ee
is positive for $\kappa_H (t) > 0$, i.e. for space-like horizons: $-2 \ge \beta < -1$, in the FRW model under consideration. In terms of the time dependence
\be
T(t) = -  \frac{1+ \beta}{4 \pi (1- t/t_s)}~,
\ee 
a rather universal time  dependence for
 $T \ge -(1 + \beta) \frac{r_b^{(1 -\beta)/\beta}}{4 \pi}$
for a fluid with boundary at $0 \le r \le r_b$  is observed (Fig.~\ref{temper}).
Note that for a stationary black hole $T=\frac{1}{8 \pi M}$ which follows from
 Eq.~(\ref{kapp3})  when $m =  M = const$ 
\cite{Hawking,Hawking:1974rv,Hawking:1974sw}.

A convenient and transparent way to present the
different properties of the solution depending on  the parameter $\beta$ is obtained by introducing the conformal time $\eta$,
\be
\eta= \int^t_0 \frac{dt'}{a(t')} =\int^{a(t)}_{a(0)} \frac{da}{a \,\dot{a}}~,
\ee
which written down explicitly  becomes
\be 
\eta = -\frac{1}{\beta}[1 - (1 - t/t_s)^{\frac{\beta}{\beta -1}}]~.
\ee

\begin{figure}
\begin{center}
\includegraphics[scale=0.4]{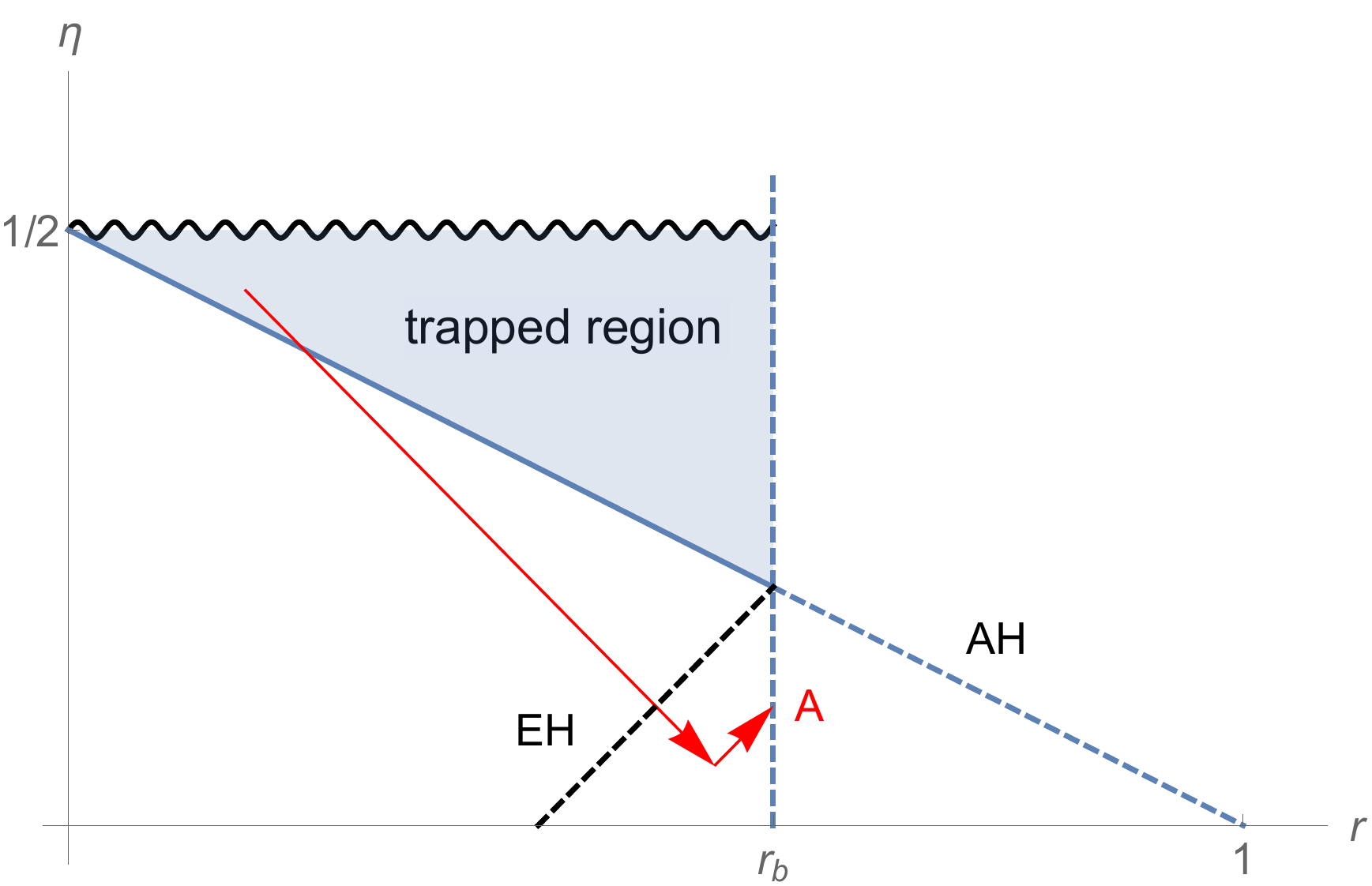}$\;\;\;\;$ \includegraphics[scale=0.4]{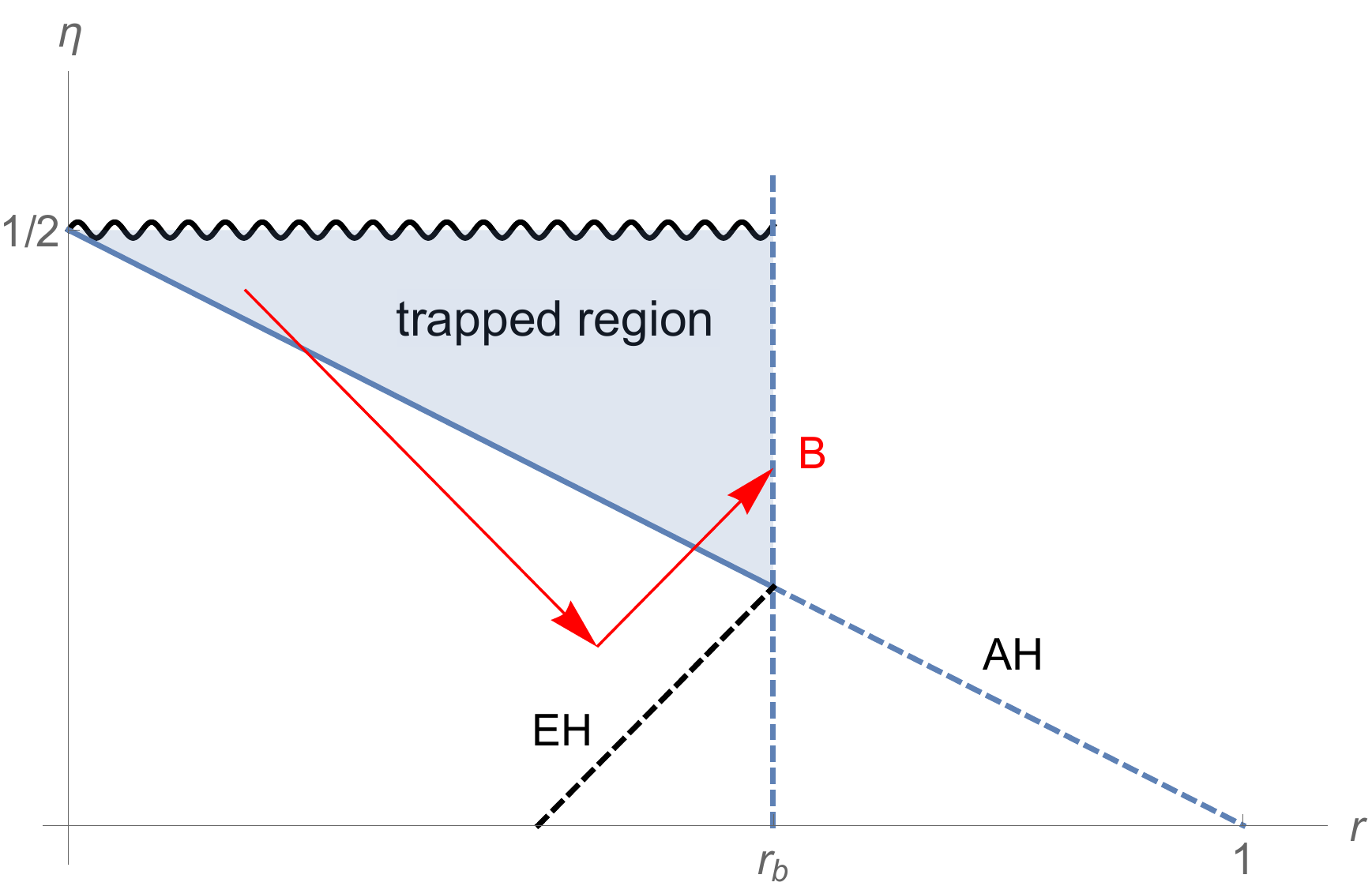}
\caption {\label{etaf}
($\eta,r$) diagram inside the stiff fluid $r \le r_b$ for a 
space-like horizon ($\beta = -2$). Left: tunneling  type-I path through the apparent  (AH) and through the
  event horizon (EH). Right: tunneling type-I path
 only through the apparent horizon, therefore the radiation is not observable outside the event horizon.}
 \end{center}
 \end{figure}

\noindent
The $(\eta,r)$ diagram is shown for the stiff fluid with $\beta = -2$ in 
Fig.~\ref{etaf}
with
\be
\eta = \frac{1}{2} [ 1 - (1 -t/t_s)^{2/3}],
\ee
i.e. $\eta = 1/2$  at $ t_s = 1/3$ and $\eta = 0$ at $t = 0$.

The apparent horizon formed during the collapse of the fluid sphere $r\le r_b$
 has its boundary at  $a^\beta_{AH} = \frac{1}{r_{AH}}$ leading to 
\be
\eta_{AH} = - \frac{1}{\beta} (1 - r) = \frac{1}{2} ( 1 -r)~.
\label{34}
\ee

The  event horizon inside the fluid boundary is determined by a radial light ray emitted at $r=0$ which just reaches the surface at $r=r_b$ at the same position as the apparent horizon
 
\be
\eta_{EH} = 
r + \frac{1}{\vert \beta \vert} (1 - r_b) - r_b~ = r + \frac{1}{2} - \frac{3r_b}{2}~.
\label{35}
\ee

\begin{figure}
\begin{center}
\includegraphics[scale=0.4]{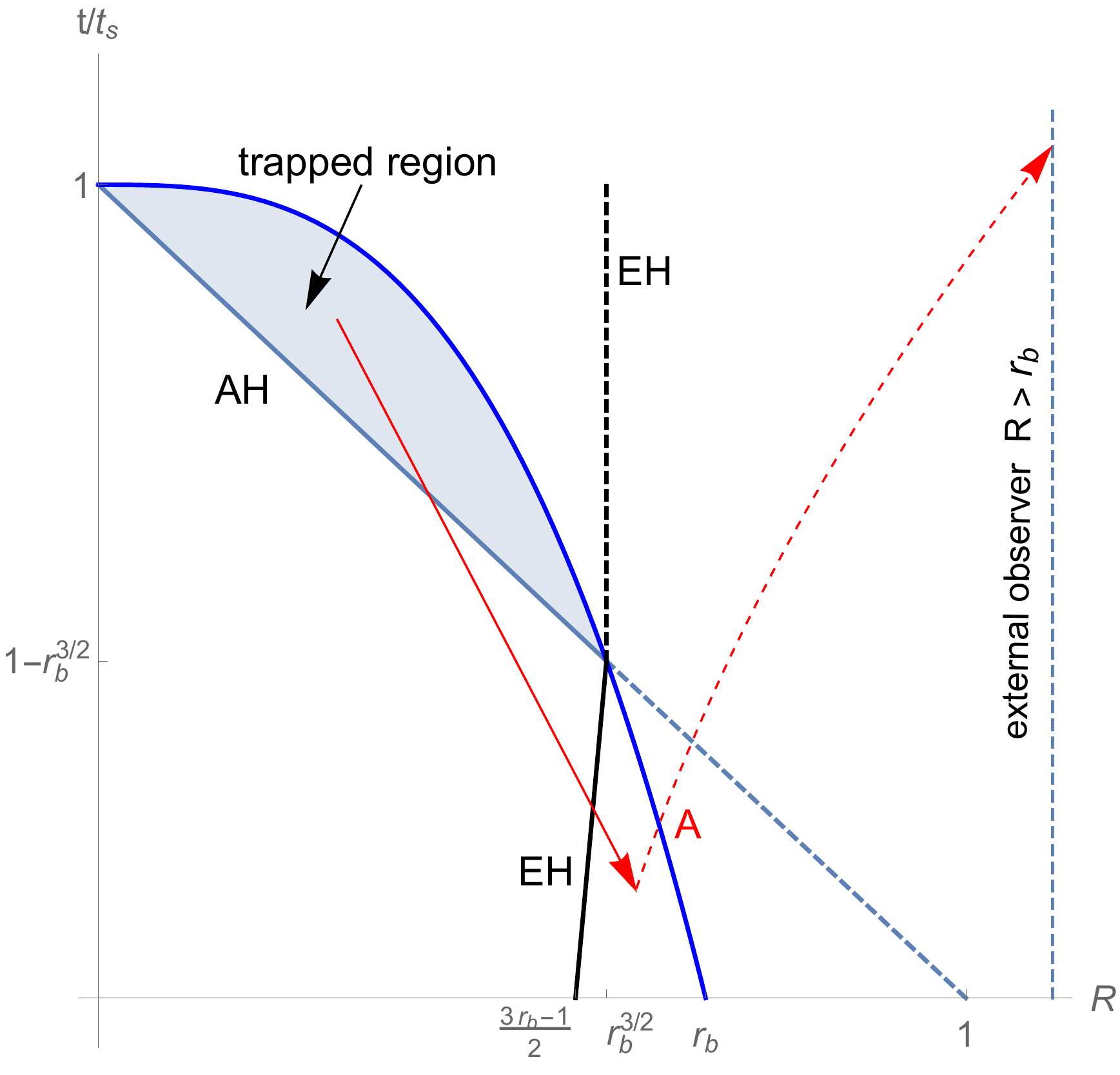}
\caption {\label{stiff} ($t, R$) diagram with space-like horizon ($\beta = -2$) and tunneling
type-I  path (red) crossing  the apparent (AH) and the event horizon (EH), finally reaching an external observer.}
\end{center}
\end{figure}

In Fig.~\ref{etaf} (left) we indicate a possible  tunnelling path through the apparent horizon AH of the stiff fluid: the type-I path $\Delta r = - \Delta \eta > 0$ starts in the trapped region, but crosses also the  event horizon EH
for $r \le r_b$, outside the trapped region, 
 before it reaches the particle-antiparticle vertex. Finally it crosses the boundary of the fluid at point A, which is outside the event horizon,   ending at an outside observer; see also Fig.~\ref{stiff}
 (c.f. Fig.~4 in \cite{Vanzo:2011wq}).  
In Fig.~\ref{etaf} (right) the pair creation vertex is inside the  EH,
therefore the radiation is not observable for an outside observer.

In order to obtain the full spacetime the metric inside the collapsing
fluid has to be matched with the outer region. 
We  perform the matching by using the Schwarzschild metric.

Outside the collapsing  fluid we choose the Schwarzschild metric
\be
ds^2 = -f (R) dT^2 + \frac{dR^2}{f(R)} + R^2 d \Omega_2^2, ~~f(R) = 1 - \frac{2M}{R},
\label{37}
\ee
which has to be matched with the interior FRW metric (\ref{FW}) at the boundary of a spherical junction hypersurface $r=r_b$. 
 From the angular parts of the two metrics, we can identify at the boundary 
\be
R = r_b a(t) \equiv R_b(t),
\label{40}
\ee
where $t$ is the proper time of the collapsing fluid and $0 \leq R_b \leq r_b$.
From the Israel junction conditions, the Misner-Sharp mass (\ref{32}) at the boundary is $2 m(r_b, t) =R_b(t)(1-f(R_b (t)))$, 
which then gives
\be
2M (t) =\frac{R^{2\beta+1}_b (t)}{r_b^{2(\beta-1)}},
\label{36}
\ee
leading to a time-dependent mass $M(t)$.   Note that   $2M (t = 0) = r^3_b$.

Following the treatment of gravitational collapse of uniform
perfect fluids as described in \cite{Adler:2005vn}, the Painlev\'e-Gullstrand coordinate system \cite{Poisson:2004} is well suited  for  the  collapse under consideration.
For the interior spacetime  the radial coordinate R,  Eq.~(\ref{radR}),
is introduced, leading  to Eq.~(\ref{3a})
with
\be
\psi (R,t) =  R  H < 0.
\label{A3}
\ee

\begin{figure}
\begin{center}
\includegraphics[scale=0.6]{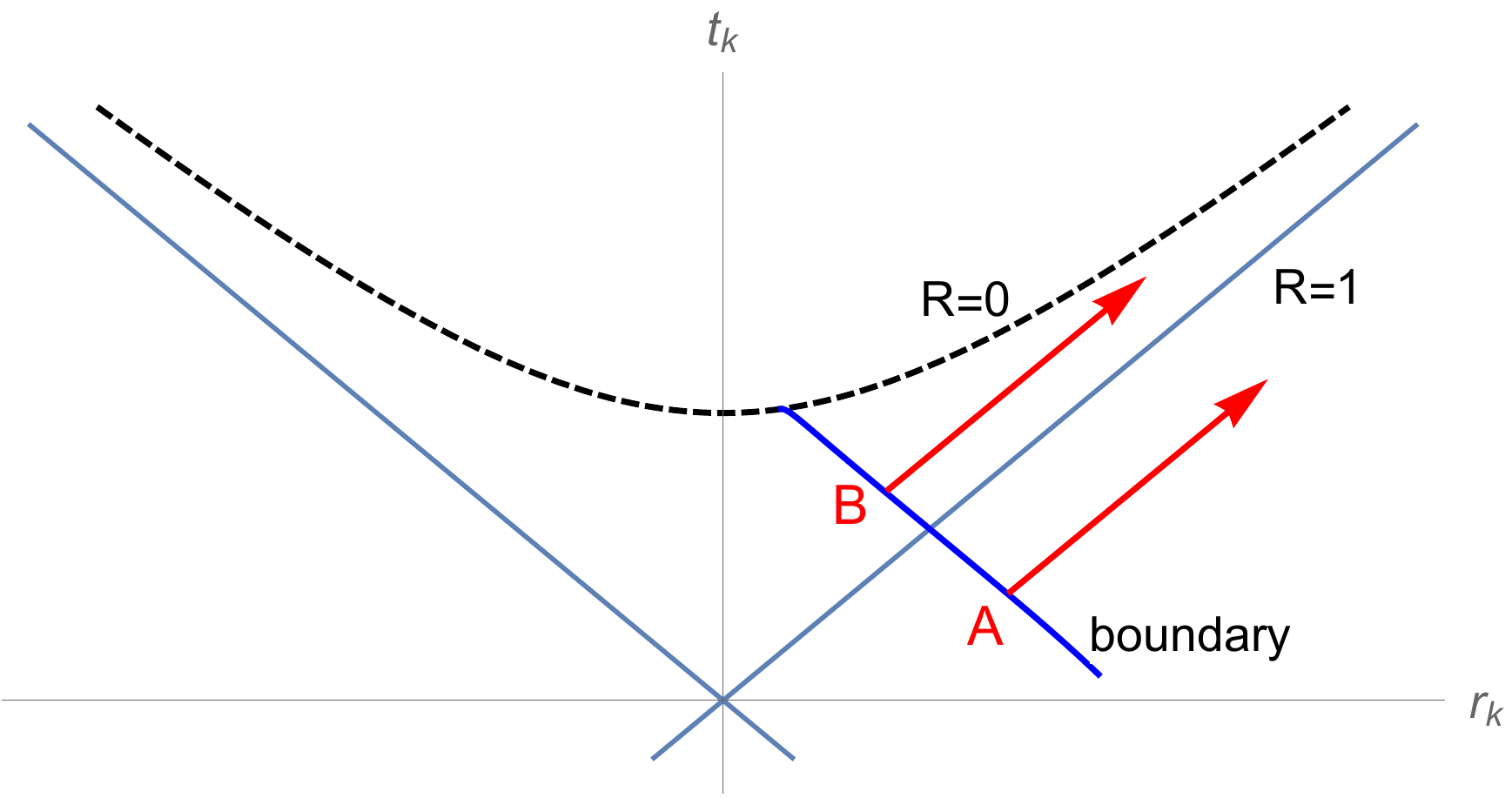}
\caption{ \label{krusk}Space-like horizon ($\beta = -2$): Kruskal coordinates outside
 the stiff fluid. The radiation is indicated by the red arrows: only the one from A reaches the external observer at infinity.}
 \end{center}
 \end{figure}

The exterior region with the metric (\ref{37})  can be brought into the form of Painlev\'e-Gullstrand (\ref{2a}) by  the transformation
\be
T = t + g (R),
\label{A5}
\ee
where 
\be
\frac{dg}{dR} =  \frac{\psi}{1 - \psi^2}, ~~\mbox{with}~~\psi^2 = \frac{2M (t)}{R}.
\label{A7}
\ee
\noindent
In the following we discuss the case of the stiff fluid ($\beta = - 2$).
Introducing
\be
R^* = \int \frac{dR}{f(R)}~,
\ee
one derives after rescaling $R$ and $T$ by $r_b^{3/2}$:
\be 
R^* + T = R + t - \arctan R~,
\ee
\be
R^* - T = R - t - \frac{1}{2} \ln \vert \frac{1 + R}{1 - R} \vert~.
\ee
Finally, the Kruskal coordinates \cite{Hamilton} read
\be 
r_K + t_K = \exp[(R^* + T)/2] ~,
\ee 
and 
\be
\pm (r_K - t_K) = \exp[(R^* - T)/2] ~,
\ee
depending on $R > 1$ or $R < 1$, resp.
The corresponding diagram is shown in Fig.~\ref{krusk}.

\subsection{Time-like horizon}

In case of a time-like horizon radiation is possible, as seen as follows:
one takes a path with decreasing radial coordinate $\Delta r < 0$,
i.e. $\int_{\gamma} = -  {\int \hspace*{-0.5cm} \searrow}$ in Eq.~(\ref{intsea})
to obtain
\be
\Im  m I =  -\frac{1}{\kappa_H} 
 \Im m   {\int \hspace*{-0.5cm} \searrow}
  \frac{\omega}{r - r_H -i\epsilon} dr =
 -\frac{\pi \omega}{\kappa_H} ~,
\ee
such that the Hawking temperature becomes
\be
 T= - \frac{\kappa_H}{2 \pi}~,
\ee
which is positive for $\kappa_H (t) < 0$, i.e. for time-like horizons:
 $- 1 < \beta < 0$, in the FRW model under consideration. 

\noindent
For the typical case of the OS-model \cite{Oppenheimer:1939ue},
with $\beta = - 1/2$, a  tunnelling path (type-II)
with the vertex inside the trapped region and crossing the time-like horizon is possible, but the radiation ends at the center $r=0$. For the interior and exterior geometry of the collapsing sphere of dust see Fig.~7 
  in \cite{Hartle}.


This case of Hawking radiation during the collapse of the fluid 
in the presence of a time-like dynamical horizon is trapped, see also
 the discussion in  \cite{Ellis:2013oka,Ellis:2014jja}: it
 does not end up at infinity.

\section{Discussion}

In summary  to obtain   the radiation temperature $T_H$ 
in the presence of a space-like or  a time-like trapped horizon  one may just take the absolute value in terms of the surface gravity on the horizon,
\be
T_H = \frac{\vert \kappa_H \vert}{2 \pi}~.
\ee
Concerning the regularization step one directly obtains this result
- instead of using Eq.~(\ref{intsea}) -  
by considering as example the integral of an   arbitrary function $ f(r) \approx \kappa_H (r-r_0)$ near the pole $r=r_0$,  where   $\kappa_H$ does not depend on $r$, 
\be
I = \Im m {\int \hspace*{-0.5cm} \searrow}
 \frac{dr}{f(r) - i\epsilon} = 
\ee
\be
  = \pi {\int \hspace*{-0.5cm} \searrow} \delta [f(r)] dr
 = \frac{\pi}{\vert \partial_r f(r_0)\vert} =   \frac{\pi}{\vert \kappa_H \vert}, 
\ee
in terms of the absolute value of  $\kappa_H$.

However, this consideration alone do not contain information if the Hawking radiation can reach an asymptotic observer. In order to decide this a more thorough investigation has to be made by checking  if the Hawking radiation crosses both, the apparent and event horizons. 
In this note we reanalyzed the case of a collapsing fluid in a flat FRW background where we showed that only Hawking radiation from  space-like trapping horizons can reach an external observer at infinity.

\vspace{0.6cm}

{\it{Acknowledgements}}
We thank L. Vanzo for useful comments and A.B. Nielsen for
 discussions on earlier versions of this manuscript. S. Stricker was supported by the Austrian science Fund (FWF) under project P26328.


\begin{thebibliography}{0}%
\makeatletter
\providecommand \@ifxundefined [1]{%
 \@ifx{#1\undefined}
}%
\providecommand \@ifnum [1]{%
 \ifnum #1\expandafter \@firstoftwo
 \else \expandafter \@secondoftwo
 \fi
}%
\providecommand \@ifx [1]{%
 \ifx #1\expandafter \@firstoftwo
 \else \expandafter \@secondoftwo
 \fi
}%
\providecommand \natexlab [1]{#1}%
\providecommand \enquote  [1]{``#1''}%
\providecommand \bibnamefont  [1]{#1}%
\providecommand \bibfnamefont [1]{#1}%
\providecommand \citenamefont [1]{#1}%
\providecommand \href@noop [0]{\@secondoftwo}%
\providecommand \href [0]{\begingroup \@sanitize@url \@href}%
\providecommand \@href[1]{\@@startlink{#1}\@@href}%
\providecommand \@@href[1]{\endgroup#1\@@endlink}%
\providecommand \@sanitize@url [0]{\catcode `\\12\catcode `\$12\catcode
  `\&12\catcode `\#12\catcode `\^12\catcode `\_12\catcode `\%12\relax}%
\providecommand \@@startlink[1]{}%
\providecommand \@@endlink[0]{}%
\providecommand \url  [0]{\begingroup\@sanitize@url \@url }%
\providecommand \@url [1]{\endgroup\@href {#1}{\urlprefix }}%
\providecommand \urlprefix  [0]{URL }%
\providecommand \Eprint [0]{\href }%
\providecommand \doibase [0]{http://dx.doi.org/}%
\providecommand \selectlanguage [0]{\@gobble}%
\providecommand \bibinfo  [0]{\@secondoftwo}%
\providecommand \bibfield  [0]{\@secondoftwo}%
\providecommand \translation [1]{[#1]}%
\providecommand \BibitemOpen [0]{}%
\providecommand \bibitemStop [0]{}%
\providecommand \bibitemNoStop [0]{.\EOS\space}%
\providecommand \EOS [0]{\spacefactor3000\relax}%
\providecommand \BibitemShut  [1]{\csname bibitem#1\endcsname}%
\let\auto@bib@innerbib\@empty
\end{thebibliography}%


\vspace{0.6cm}

\begin{thebibliography}{99}



\bibitem{Parikh:1999mf}
  M.~K.~Parikh and F.~Wilczek,
  Phys.\ Rev.\ Lett.\  {\bf 85} (2000) 5042
  [hep-th/9907001].

\bibitem{Visser:2001kq}
  M.~Visser,
  ``Essential and inessential features of Hawking radiation,''
  Int.\ J.\ Mod.\ Phys.\ D {\bf 12} (2003) 649
  [hep-th/0106111].

\bibitem{Vanzo:2011wq}
for a review and references:  

L.~Vanzo, G.~Acquaviva and R.~Di Criscienzo,
  ``Tunnelling Methods and Hawking's radiation: achievements and prospects,''
  Class.\ Quant.\ Grav.\  {\bf 28} (2011) 183001
  [arXiv:1106.4153 [gr-qc]].

\bibitem{Vanzo:2008uq}
  L.~Vanzo,
  ``Some results on dynamical black holes,''
  arXiv:0811.3532 [gr-qc].






\bibitem{Nielsen:2005af}
  A.~B.~Nielsen and M.~Visser,
  ``Production and decay of evolving horizons,''
  Class.\ Quant.\ Grav.\  {\bf 23} (2006) 4637
  [gr-qc/0510083].


\bibitem{Nielsen:2008cr}
  A.~B.~Nielsen,
  ``Black holes and black hole thermodynamics without event horizons,''
  Gen.\ Rel.\ Grav.\  {\bf 41} (2009) 1539
  [arXiv:0809.3850 [hep-th]].






\bibitem{Tian:2014sca}
  D.~W.~Tian and I.~Booth,
  ``Apparent horizon and gravitational thermodynamics of the Universe: Solutions to the temperature and entropy confusions, and extensions to modified gravity,''
  arXiv:1411.6547 [gr-qc].


\bibitem{Ashtekar:2004cn}
  A.~Ashtekar and B.~Krishnan,
  ``Isolated and dynamical horizons and their applications,''
  Living Rev.\ Rel.\  {\bf 7} (2004) 10
  [gr-qc/0407042].
  


\bibitem{Hayward:2008jq}
  S.~A.~Hayward, R.~Di Criscienzo, L.~Vanzo, M.~Nadalini and S.~Zerbini,
  ``Local Hawking temperature for dynamical black holes,''
  Class.\ Quant.\ Grav.\  {\bf 26} (2009) 062001
  [arXiv:0806.0014 [gr-qc]].

\bibitem{DiCriscienzo:2010zza}
  R.~Di Criscienzo, S.~A.~Hayward, M.~Nadalini, L.~Vanzo and S.~Zerbini,
  ``Hamilton-Jacobi tunneling method for dynamical horizons in different coordinate gauges,''
  Class.\ Quant.\ Grav.\  {\bf 27} (2010) 015006.

\bibitem{Senovilla:2014ika}
  J.~M.~M.~Senovilla and R.~Torres,
  ``Particle production from marginally trapped surfaces of general spacetimes,''
  arXiv:1409.6044 [gr-qc].

\bibitem{Nielsen:2008dj}
  A.~B.~Nielsen,
  ``Black Holes without Event Horizons,''
  J.\ Korean Phys.\ Soc.\  {\bf 54} (2009) 2576
  [arXiv:0802.3422 [gr-qc]].
  
  
\bibitem{Baier:2014ita}
  R.~Baier, H.~Nishimura and S.~A.~Stricker,
  ``Scalar field collapse with negative cosmological constant,''
 Class.\ Quant.\ Grav.\  {\bf 32} (2015) 135021
  [arXiv:1410.3495 [gr-qc]].



\bibitem{Joshi:2002}
  P.~S.~Joshi, ``Global Aspects in Gravitation and Cosmology,''
Oxford Univ. Press 2002.

 
\bibitem{Joshi:2011hb}
  P.~S.~Joshi,
  ``Key problems in black hole physics today,''
 Bull. Astr. Soc. India {\bf 39} (2011) 1, 
  arXiv:1104.3741 [gr-qc].

\bibitem{Joshi:2008zz}
  P.~S.~Joshi,
  ``Gravitational Collapse and Spacetime Singularities,''
Cambridge University Press, 2007.
  
 
\bibitem{Joshi:2007zza}
  P.~S.~Joshi and R.~Goswami,
  ``On trapped surface formation in gravitational collapse,''
  Class.\ Quant.\ Grav.\  {\bf 24} (2007) 2917.

\bibitem{Hayrev}
 ``Black Holes - New Horizons,''
ed. S.~A.~Hayward, World Scientific, 2013.  

\bibitem{Oppenheimer:1939ue}
  J.~R.~Oppenheimer and H.~Snyder,
  ``On continued gravitational contraction,''
  Phys.\ Rev.\  {\bf 56} (1939) 455.
  
 

\bibitem{Adler:2005vn}
  R.~J.~Adler, J.~D.~Bjorken, P.~Chen and J.~S.~Liu,
  ``Simple analytic models of gravitational collapse,''
  Am.\ J.\ Phys.\  {\bf 73} (2005) 1148
  [gr-qc/0502040].
  

\bibitem{Hayward:1994}
  S.~A.~Hayward,
  ``General laws of black-hole dynamics,''
  Phys.\ Rev.\ D {\bf 49} (1994) 6467.
     
\bibitem{Hayward:1996}
  S.~A.~Hayward,
  ``Gravitational energy in spherical symmetry,''
  Phys.\ Rev.\ D {\bf 53} (1996) 1938.

\bibitem{Hayward:2000ca}
  S.~A.~Hayward,
  ``Black holes: New horizons,''
  [gr-qc/0008071].
  
\bibitem{Booth:2005qc}
  I.~Booth,
  ``Black hole boundaries,''
  Can.\ J.\ Phys.\  {\bf 83} (2005) 1073
  [gr-qc/0508107].



\bibitem{Booth:2005ng}
  I.~Booth, L.~Brits, J.~A.~Gonzalez and C.~Van Den Broeck,
  ``Marginally trapped tubes and dynamical horizons,''
  Class.\ Quant.\ Grav.\  {\bf 23} (2006) 413
  [gr-qc/0506119].
  
\bibitem{Poisson:2004}
  E.~Poisson,
  ``A Relativist's Toolkit: The Mathematics of Black-Hole Mechanics,''
  Cambridge Univ. Press, 2004.
  
\bibitem{Misner:1964je}
  C.~W.~Misner and D.~H.~Sharp,
  ``Relativistic equations for adiabatic, spherically symmetric gravitational collapse,''
  Phys.\ Rev.\  {\bf 136} (1964) B571.


\bibitem{Hayward:1997jp}
  S.~A.~Hayward,
  ``Unified first law of black hole dynamics and relativistic thermodynamics,''
  Class.\ Quant.\ Grav.\  {\bf 15} (1998) 3147
  [gr-qc/9710089].

\bibitem{Pielahn:2011ra}
  M.~Pielahn, G.~Kunstatter and A.~B.~Nielsen,
  ``Dynamical Surface Gravity in Spherically Symmetric Black Hole Formation,''
  Phys.\ Rev.\ D {\bf 84} (2011) 104008
  [arXiv:1103.0750 [gr-qc]].


\bibitem{Cai:2008gw}
  R.~G.~Cai, L.~M.~Cao and Y.~P.~Hu,
  ``Hawking Radiation of Apparent Horizon in a FRW Universe,''
  Class.\ Quant.\ Grav.\  {\bf 26} (2009) 155018
  [arXiv:0809.1554 [hep-th]].


\bibitem{Hawking}
St. ~Hawking and R.~Penrose,
``The nature of space and time,''
Princeton Univ. Press, 1996.

\bibitem{Hawking:1974rv}
  S.~W.~Hawking,
  ``Black hole explosions,''
  Nature {\bf 248} (1974) 30.

\bibitem{Hawking:1974sw}
  S.~W.~Hawking,
  ``Particle Creation by Black Holes,''
  Commun.\ Math.\ Phys.\  {\bf 43} (1975) 199
   [Erratum-ibid.\  {\bf 46} (1976) 206].

\bibitem{Hamilton} 
Andrew J.~S.~Hamilton,
" General Relativity, Black Holes, and Cosmology", proto-book, 2014.


\bibitem{Hartle}
J.~B.~Hartle, 
"Relativistic stars, gravitational collapse, and black holes", in
Relativity, Astrophysics and Cosmology, ed. W.~Israel (Dordrecht, Reidel) p. 153, 1973.



\bibitem{Ellis:2013oka}
  G.~F.~R.~Ellis,
  ``Astrophysical black holes may radiate, but they do not evaporate,''
  arXiv:1310.4771 [gr-qc].





\bibitem{Ellis:2014jja}
  G.~F.~R.~Ellis, R.~Goswami, A.~I.~M.~Hamid and S.~D.~Maharaj,
  ``Astrophysical Black Hole horizons in a cosmological context: Nature and possible consequences on Hawking Radiation,''
  Phys.\ Rev.\ D {\bf 90} (2014) 8,  084013
  [arXiv:1407.3577 [gr-qc]; and references therein.


\end{thebibliography}
\end{document}